\documentclass[prb,aps,showpacs,two column]{revtex4}

\usepackage{graphicx}
\usepackage{epsfig}
\usepackage{amsmath}
\usepackage{amssymb}
\usepackage{amsthm}
\usepackage{bm}
\def\topfraction{1}
\def\bottomfraction{1}
\def\textfraction{0.05}
\def\floatpagefraction{0.95}
\begin{document}

\def\topfraction{1}
\def\bottomfraction{1}
\def\textfraction{0.05}
\def\floatpagefraction{0.95}
\textfloatsep = 0.5cm \floatsep = 0.0cm \setcounter{topnumber}{2}
\setcounter{bottomnumber}{2}

\title{Second Harmonic Generation in Gapped Graphene}


\author{ Godfrey Gumbs}

 \affiliation{Department of Physics and Astronomy, Hunter College of the City University of New York,\\
695 Park Avenue, New York, NY 10065, USA}

\author{Yonatan Abranyos}

\affiliation{Department of Physics and Astronomy, Hunter College of the City University of New York,\\
695 Park Avenue, New York, NY 10065, USA}

\author{Upali Aparajita}

\affiliation{Department of Physics and Astronomy, Hunter College of the City University of New York,\\
695 Park Avenue, New York, NY 10065, USA}

\author{Oleksiy Roslyak}

\affiliation{Los Alamos National Laboratory, Los Alamos, NM 87545, USA}

\date{\today}

\begin{abstract}
The second-order nonlinear optical susceptibility $\Pi^{(2)}$ for second harmonic
generation is calculated for gapped graphene. 
The linear and second-order nonlinear plasmon excitations are investigated in context of second harmonic generation (SHG).
We report a red shift and an order of magnitude enhancement of the SHG resonance
with growing gap, or alternatively, reduced electro-chemical potential.
\end{abstract}


\maketitle

\section{Introduction}

Since the discovery of second harmonic generation (SHG) by Franken
et al. and the demonstration of the first working
laser by Maiman in  early 60-x, various nonlinear optical techniques has received considerable attention\cite{boyd}. 
At the hear of those techniques lays the response to $n-$power of the optical filed $\Pi^{(n)}$,
which , in essence, is the multi-point correlation function between the electrons of the probed substance\cite{mukamel}.
For instance, $\Pi^{(2)}$ describes various two-wave mixing such as SHG, sum and difference frequency generation (SFG,DFG) and linear
electro-optical effects (Pockets).
Those are of great importance in areas of integrated optics and optical communication, SFG based 
frequency-tunable visible lasers and DFG based optical parametric oscillators \cite{boyd}.
Typical value of $\Pi^{(2)}$ is of the order of $\sim\, 1.67\times 10^{-12}  m/V$. 
Various groups \cite{1467,khurgin,rosencher,yoo,harshman,shaw,seto,cai} demonstrated substantial\footnote{Two orders of magnitude} enhancement of $\Pi^{(2)}$ for an asymmetric quantum well (QW), 
asymmetric double quantum well (DQW) and several bond-altering dipolar structures. 
In addition, there are quite a few papers\cite{kuwatsuka,tsang,fejer,west} dealing with the calculation of $\Pi^{(2)}$ for a single QW biased by an electric field. 
\par
\begin{figure}[htbp]
\centering
\includegraphics[width=0.4\textwidth]{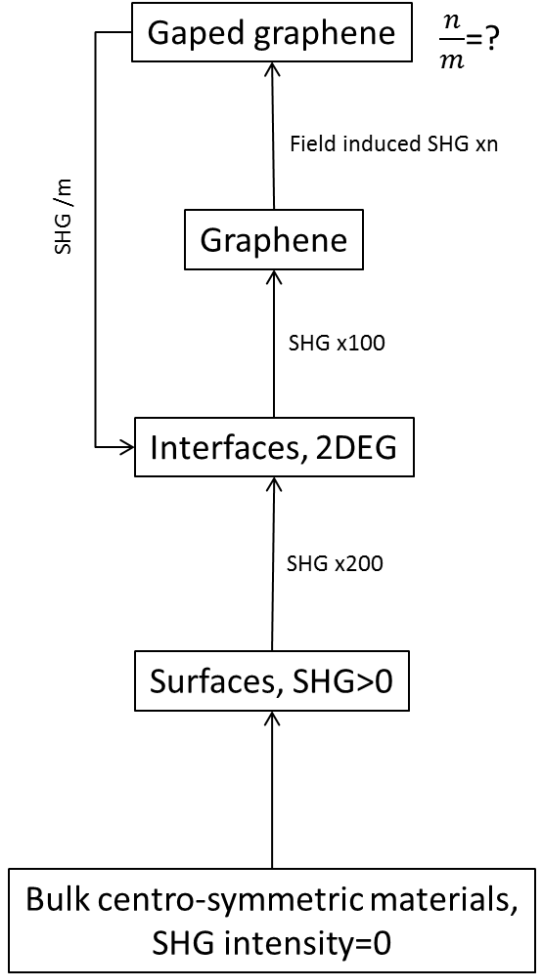}
\caption{(Color online) Hierarchy of SHG enhancement. }
\label{FIG:1}
\end{figure}

SHG is a powerful optical tool for probing surfaces, thin films\cite{shen}, multilayer graphene\cite{bykov} as well as hetero interfaces such as two dimensional electron gas\cite{stern}  of centrosymmetric materials.
 In the dipole approximation, SHG is prohibited in the bulk of such materials, while at surfaces and interfaces the central symmetry is broken. 
For the two dimensional electron gas, SHG gives two orders of magnitude larger signal when compared with surfaces. 
Recently,  an additional two orders of magnitude enhancement of the SHG  signal in graphene compared with $GaAs$ two dimensional electron gas was predicted\cite{mikhailov}, 
as shown in Fig.\ref{FIG:1}. 
The author also reported an order of magnitude larger linear response in that system.

\par
A typical graphene-based SHG experimental set-up involves  specular light reflection in
the wave length range of $730-830 \; nm$. Reflected SHG radiation is spectrally selected and
quadratic dependence of the signal on the incoming pulse intensity must be assured\cite{bykov}.
The inversion symmetry between A and B sub-lattices in graphene can be broken by external fields
causing so-called field induced SHG\cite{pan}. On the level of graphene electronic spectra, the
external influence opens up a gap. Examples of such Dirac cone perturbation are  multilayer
epitaxially grown graphene\cite{ohta}, circularly polarized light\cite{kibis} and underlying
substrate\cite{giovannetti}. On one hand, the gap makes graphene   behave more like conventional
2DEG thus lowering the SHG intensity. On the other hand, the field induced SHG boosts   the signal.
In this paper we investigate the interplay between these two  effects schematically
as shown in Fig.\ref{FIG:2}.

\par
Our paper focuses on gapped
graphene. 
As will be discussed later, the gap in the graphene electronic spectrum means
broken inversion symmetry, thereby promising enhanced second-order response.
The resonances in linear density-density response are known as plasmons.
We shall demonstrate the existence of similar plasmon-like resonances in the second-order
response, in particular the part corresponding to SHG.

\section{Model for Gapped Graphene }

In the low energy regime near the Dirac points, the electronic spectrum
of graphene exhibits the familiar linear dispersion with zero energy gap at
the two Dirac points ($\mathbf{K},\mathbf{K}^\prime$). Opening a gap in the spectrum of
graphene generally involves breaking  the underlying inversion symmetry. There are
several ways in which  the symmetry might be broken. These include  coupling with a quantized
circularly polarized field,  breaking of the sub-lattice symmetry, spin-orbit
coupling via the Rashba interaction, reduction in dimension leading edge
effects in  zigzag nano-ribbons or confinement in armchair nano-ribbons.
For small  deviation $\mathbf{k}$ in  the electron momentum from the Dirac points,
 the tight-binding model reduces to the eigenvalue equation
$H_g \vert{\lambda}\rangle= { E}  \vert{\lambda}\rangle$, where the Hamiltonian is given by
\begin{figure}[htbp]
\centering
\includegraphics[width=0.5\textwidth]{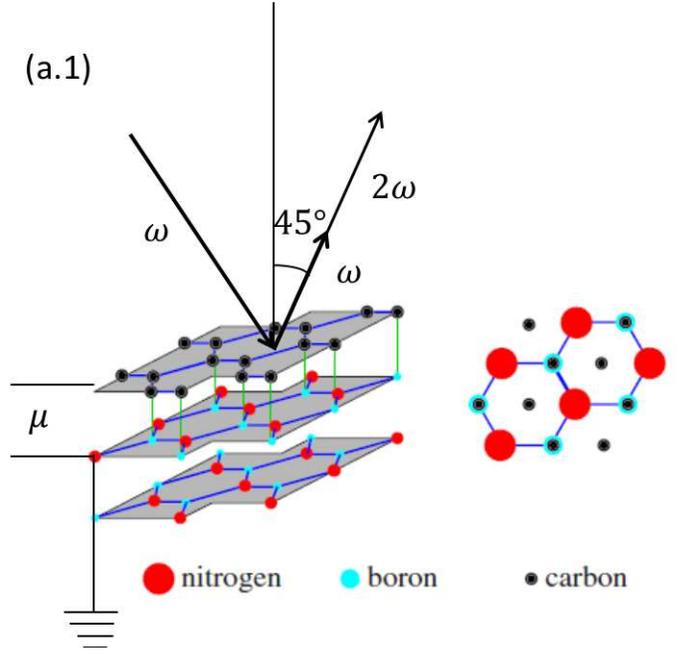}
\caption{(Color online) Panel (a.1) schematic of the SHG specular reflection experiment}
\label{FIG:2}
\end{figure}
\begin{equation}
\label{EQ:HAMILTONIAN}
H_g=\left( \begin{array}{cccc}
E_{g}/2 & \hbar v_Fk^* & 0 & 0 \\
\hbar v_Fk& -E_{g}/2 & 0 & 0\\
0 & 0 & -E_{g}/2 & -\hbar v_Fk^*\\
0 & 0 & -\hbar v_Fk & E_{g}/2\end{array} \right)\ 
\end{equation}
where $k=k_x+ik_y$ is the complex wave-vector, $v_F$ is the Fermi  velocity 
and the electronic states for the  A,B sublatices are $ \langle \mathbf{k}
\vert{\lambda}\rangle=[\psi_A({\bf k}),\psi_B({\bf k}),-\psi'_A({\bf k}),-\psi'_B({\bf k})]$.
The corresponding eigenvalues yield the conduction and valence bands shown
schematically  in Fig. \ref{FIG:2}:
\par

\begin{equation}
\label{EQ:DISPERSION}
{ E}_{\pm} = \pm\sqrt{E^2_g/4+(\hbar v_fk)^2} \  
\end{equation}
Here $E_g$ is the energy gap at $k=0$.
The alternating sign of $E_g$ in Eq.\eqref{EQ:HAMILTONIAN} indicates broken symmetry
between the A and B sub-lattices. In the next section, we employ this feature in order to
generate second-order nonlinear polarization.


\section{Linear and second order-response of gapped graphene subjected
to a harmonic potential}

\begin{figure}[htbp]
\centering
\includegraphics[width=0.5\textwidth]{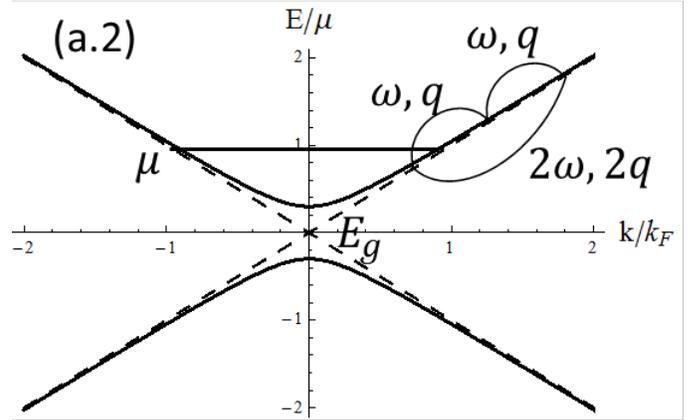}
\caption{(Color online) Intraband induced SHG in gaped graphene in the  long wavelength approximation. }
\end{figure}
We now consider the dynamics of graphene interaction with an oscillating single-mode
electromagnetic  field described by the   potential
$\Phi({\bf r},t)=\Phi_{{\bf q}\omega}e^{i({\bf q\cdot r}-\omega t)}+c.c.$.
In this section, we   derive a formal expression for the response function due to an external
perturbation   up to  second order and further process it in the long wavelength approximation.
Perturbative treatment of the density matrix suits best for that purpose.
The reduced density operator $\hat\rho$ satisfies
the equation of motion:

\begin{gather}
\label{density}
i \hbar\frac{\partial\hat\rho}{\partial t}= [H_g+H_{int},\hat\rho]\\
\notag
H_{int}=-e\Phi({\bf r},t) \ .
\end{gather}
The external field is turned on adiabatically, i.e.,

\begin{equation}
\begin{cases}
  \hat{\rho_0}|\lambda\rangle=f_{\lambda}|\lambda\rangle,  & \text{Initial condition} \\
  f_{\lambda}=1-\theta\left({E_{\lambda}-\mu}\right), & \text{Distribution at}\;\;
  \Phi_{{\bf q},\omega}=0
\end{cases}
\label{density_sol}
\end{equation}
with $\mu$ being the chemical potential. We shall seek solutions of Eqs. \eqref{density}
subjected to the conditions given in  \eqref{density_sol} in the density fluctuations form

\begin{gather}
\label{netsolution}
Tr\ [\hat{\rho}]= \rho_0+\rho_{{\bf q},
\omega}e^{i({\bf q\cdot r}-\omega t)}
\rho_{2{\bf q},2\omega}e^{2i({\bf q\cdot r}-\omega t)}+c.c.\\
\notag
\rho_{{\bf q},\omega}=-e^{2}\Pi_{{\bf q},\omega;
{\bf q},\omega}\Phi_{{\bf q},\omega}\\
\label{fluctuations}
\rho_{2{\bf q},2\omega}=+e^{3}\Pi_{2{\bf q},2\omega ;{\bf q},\omega}\Phi^2_{{\bf q},\omega} \ ,
\end{gather}
where we took into account conservation of momentum $2{\bf q=q+q}$ and energy
$2\omega =\omega+\omega$.

\par
A general formalism for calculating the  response to arbitrary order of a quantum
system that is based on Feynman-Keldysh  (FK) diagrams was developed by Mukamel\cite{mukamel}.
The linear response is given by two FK diagrams in Fig. (6.5 c) in  Ref.\ \cite{mukamel}.
Translating those diagrams into an expression for the polarization and replacing the dummy
indices of the quantum states to those composite indices of graphene as:

\begin{gather*}
\rho_0 P(a)=f_{\lambda}\\
a \to \lambda, \; b \to \lambda'
\end{gather*}
yields  the well-known Lindhard formula
\begin{figure}[h]
\centering
\includegraphics[width=0.48\textwidth]{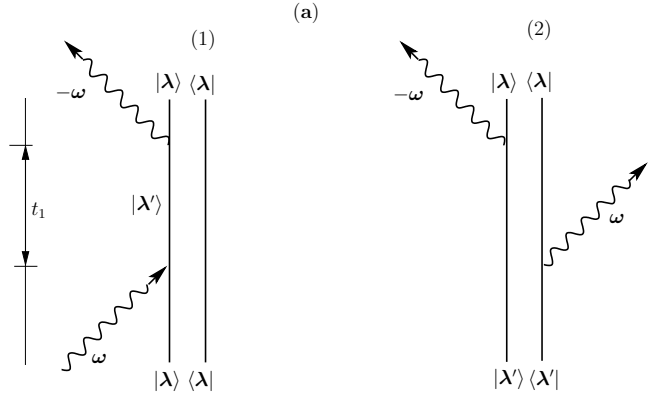}
\caption{(Color online)  Feynman diagrams used for calculating 
first-order contribution to the polarization function.}
\end{figure}
\begin{gather}
\Pi^{(1)}_{{\bf q},\omega}=\Pi_{{\bf q},\omega;
{\bf q},\omega}=-\frac{1}{\hbar}\sum_{\lambda,\lambda'}\frac{(f_{\lambda'}-f_{\lambda})|\mu_{\lambda,\lambda'}|^2}
{E_{\lambda'}-E_{\lambda}+\hbar(\omega+i\gamma)}\\
\notag
=\sum_{\lambda,\lambda'}(f_{\lambda'}-f_{\lambda})I_{\lambda\lambda'}(\omega)|\mu_{\lambda,\lambda'}|^2
\end{gather}
with   initial state $\vert{\lambda}\rangle = \vert{s,\mathbf{k}}\rangle$. Here
$s=\pm$  labels the conduction/valence bands.
The final state is $\vert{\lambda'}\rangle = \vert{s^\prime ,\mathbf{k}+\mathbf{q}}\rangle$.
The   overlap factor, given by the product of transition dipole moments, is

\begin{eqnarray*}
\mu_{\lambda,\lambda'}\mu^*_{\lambda,\lambda'}
&\approx&\langle\lambda'|e^{-i{\bf q\cdot r}}|\lambda\rangle
\langle\lambda|e^{i{\bf q\cdot r}}|\lambda'\rangle\\
&=&\frac{1}{2}\left(1+s s'\frac{\hbar^2 v^2_F{\bf k \cdot(k+q)}
+(E_g/2)^2}{E_{\bf k}E_{\bf k+q}} \right). \ 
\end{eqnarray*}

\par
\begin{figure}[h]
\centering
\includegraphics[width=0.48\textwidth, height=0.5\textwidth]{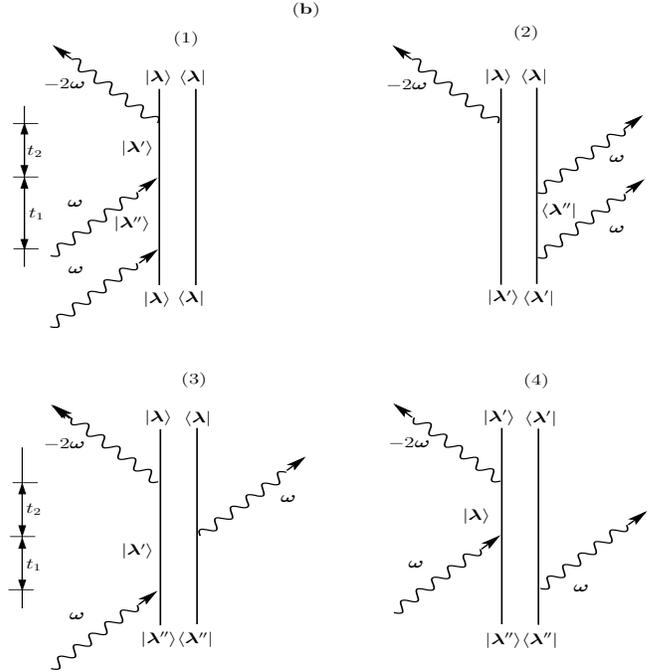}
\caption{(Color online)  Feynman diagrams used for calculating 
the second-order contributions to the polarization function.}
\label{FIG:3}
\end{figure}

The second-order response function has four Feynman diagrams,  shown in Fig.\ \ref{FIG:3}
 and Eq. (6.22) in Ref.\cite{mukamel}, thus yielding
 
\begin{eqnarray}
\Pi^{(2)}_{2{\bf q},2\omega}=\Pi_{2{\bf q},2\omega ;{\bf q},\omega}&=&\frac{1}{\hbar^2}
\sum_{a,b,c}\rho_0P(a)\Big[\mu_{ab}\mu_{bc}\mu^*_{ca}
I_{ca}(2\omega)I_{ba}\nonumber\\
&-&\mu_{ab}\mu^*_{bc}\mu_{ac}I_{bc}(2\omega)I_{ba}(\omega)\nonumber\\
&+&\mu^*_{ab}\mu_{bc}\mu_{ca}I_{ab}(2\omega)I_{ac}\nonumber\\
&-&\mu_{ab}\mu^*_{bc}\mu_{ca}I_{bc}(2\omega)I_{ac}(\omega)\Big] \ ,
\label{linear2}
\end{eqnarray}
where we have used the replacement $\left({1/\hbar}\right)^2\sum_{perm(\omega_1,\omega_2=\omega)}\to
2/\hbar^2$. Due to the fact $a,b,c$ are dummy indices, we replace

\begin{eqnarray*}
\text{First term}&&\; c\to\lambda',\;a\to\lambda,\;b\to\lambda''\\
\text{Second term}&&\; b\to\lambda',\;c\to\lambda,\;a\to\lambda''\\
\text{Third term}&&\; a\to\lambda',\;b\to\lambda,\;c\to\lambda''\\
\text{Fourth term}&&\; c\to\lambda',\;c\to\lambda,\;a\to\lambda''
\end{eqnarray*}
Here, the ``initial" state is $\vert{\lambda}\rangle=\vert{s,\mathbf{k}}\rangle$.
The doubly excited ``final" state is  $\vert{\lambda'}\rangle=\vert{s',\mathbf{k}+2 \mathbf{q}}\rangle$
and the ``intermediate" state is denoted as  
$\vert{\lambda''}\rangle=\vert{s'',\mathbf{k}+ \mathbf{q}}\rangle$.
Note that the names in parentheses are just suggestive since each of those states may
 be a ground state in our formalism. Consequently, we may take $I(2\omega)$ as a common prefactor
and obtain
\begin{eqnarray}
\Pi^{(2)}_{2{\bf q},2\omega}&=&\frac{1}{\hbar^2}
\sum_{\lambda,\lambda',\lambda''}
\frac{\mu^{\ast}_{\lambda,\lambda'}}{E_{\lambda'}-E_{\lambda}+
2\hbar(\omega+i\gamma)}\nonumber\\
&\times&\Big[\frac{\mu_{\lambda,\lambda''}\mu_{\lambda'',\lambda}(f_{\lambda'}
-f_{\lambda''})}{E_{\lambda'}-E_{\lambda''}+
\hbar(\omega+i\gamma)}\nonumber\\
&-&\frac{\mu_{\lambda,\lambda''}\mu_{\lambda'',\lambda}(f_{\lambda''}-
f_{\lambda})}{E_{\lambda'}-E_{\lambda}+\hbar(\omega+i\gamma)}\Big]
\label{linear3}
\end{eqnarray}
In Eq. (\ref{linear3}), we have introduced the matrix elements

\[
\mu^*_{\lambda',\lambda}=\langle\lambda'|e^{-2i{\bf q\cdot r}}
|\lambda\rangle \ .
\]
Owing the composite nature of $\lambda$, the outer summation over those indices converts
into an integration with the help of

\[
\sum_{\lambda,\lambda',\lambda''}\to\sum_{s,s',s''}
\frac{1}{L^2}\sum_{\bf k}\to\sum_{s,s',s''}\int kdkd\phi \ ,
\]
where $\phi$ is the angle between ${\bf k}$ and ${\bf q}$. 
Without loss of  generality, we
 may assume ${\bf q}=(q_x,0)$. Calculating such integral is a formidable task (see Refs.\
 \cite{pyatkovskiy,dassarma}). However, the  long wavelength approximation simplifies it. Formally, it is
determined by the following conditions:

\begin{equation}
\begin{cases}
q\ll k_F, & k_F= \mu/\hbar v_F\\
v_Fq\ll\omega
\end{cases}
\label{long_cond}
\end{equation}
In the  microwave and infra-red regimes, Eq. (\ref{long_cond})
restricts the  wave number $q\approx 10^6{\text cm}^{-1}$. We shall
also assume high doping; $\hbar\omega\ll\mu,\;\mu\ll T$.
Under this condition, we may  neglect the inter-band transition  contributions
to the polarization since $\lim \limits_{\mathbf{q}\to 0}\mu_{\lambda,\lambda'}=\delta_{s,s'}+O(q^2)$.
Secondly, we neglect the imaginary part of the polarization function. This is
the condition necessary for undamped plasmon resonances in the region of interest.
Those facts are known from the full version of calculated linear
polarizations\cite{pyatkovskiy,dassarma}. We shall extrapolate this assumption to
$\Pi^{(2)}_{2{\bf q},2\omega}$. To proceed further, we employ the identity
$\partial E_{\lambda}/\partial q_{\alpha} = \partial E_{\lambda}/\partial k_{\alpha}$,
with $\alpha= x,y$. This, in turn, leads to the  identity

\begin{eqnarray*}
\frac{\partial f_{\lambda}}
{\partial q_{\alpha}}{\bigg|}_{q_{\alpha=0}}&=&
\frac{\partial f}
{\partial E}\frac{\partial E}
{\partial q_{\alpha}}\\
&=&\frac{\partial f}{\partial E}\frac{\partial E}
{\partial k_{\alpha}}=\frac{\partial f}
{\partial k_{\alpha}}{\bigg|}_{q_{\alpha=0}}.
\end{eqnarray*}
At zero temperature, we  keep only the linear term after expanding in powers of
$q_{\alpha}$ and we obtain

\begin{equation}
f_{\lambda'}-f_{\lambda}\approx\sum_{\alpha}
-q_{\alpha}\frac{\partial f_{\lambda}}
{\partial k_{\alpha}}{\bigg|}_{q_{\alpha=0}}=
\sum_{\alpha}q_{\alpha}\frac{\partial E_{\lambda}}{\partial k_{\alpha}} \delta(E_{\lambda}-\mu) \ .
\label{long2}
\end{equation}
Bearing in mind that the imaginary part of the polarization is zero in the region that we
are  interested in, we obtain

\begin{eqnarray}
\frac{1}{E_{\lambda'}-E_{\lambda}+\hbar(\omega
+i\gamma)}&\approx&\frac{E_{\lambda'}-E_{\lambda}}
{(\hbar\omega)^2+(E_{\lambda'}-E_{\lambda})^2}\nonumber\\
&\approx&\frac{1}{(\hbar \omega)^2}\sum_{\beta}q_{\beta}
\frac{\partial E_{\lambda}}{\partial k_{\beta}} \ .
\label{long3}
\end{eqnarray}

Substituting Eqs. (\ref{long2}) and (\ref{long3}) into
Eq. (\ref{linear3}), we get

\begin{equation}
\Pi^{(1)}_{{\bf q},\omega}\approx\sum_{\lambda}\sum_{\alpha,\beta}
\frac{q_{\alpha}q_{\beta}}{(\hbar\omega)^2}
\frac{\partial E_{\lambda}}{\partial k_{\alpha}}
\frac{\partial E_{\lambda}}{\partial k_{\beta}}
\delta\left({E_{\lambda}-\mu}\right) \ .
\label{linearlong1}
\end{equation}
In a similar way,  we obtain the SHG polarization function as

\begin{eqnarray}
\Pi^{(2)}_{2{\bf q},2\omega}&=&-\frac{3}{2}
\sum_{\lambda}\sum_{\alpha,\beta,\gamma,\delta}
\frac{q_{\alpha}q_{\beta}q_{\gamma}q_{\delta}}{(\hbar\omega)^4}\nonumber\\
&\times&\frac{\partial E_{\lambda}}{\partial k_{\alpha}}
\frac{\partial E_{\lambda}}{\partial k_{\beta}}
\frac{\partial^2 E_{\lambda}}{\partial k_{\gamma}\partial k_{\delta}}
\delta\left({E_{\lambda}-\mu}\right)\ .
\label{longnonlinear}
\end{eqnarray}
The factor of three-half in the above expression arises from the identity
$\partial E_{\lambda'}/\partial q_{\alpha} = \frac{1}{2} \partial E_{\lambda}/\partial k_{\alpha}$,
since $\vert{\lambda'}\rangle = \vert{s', \mathbf{k}+2 \mathbf{q}}\rangle$.
The general form of Eqs. \eqref{linearlong1}, \eqref{longnonlinear} were obtained in
Ref. \cite{vafek,mikhailov}. Their adaptation to our case requires the following set of expressions:

\begin{gather}
\label{EQ:TRANSFORM}
\delta(E_{\lambda}-\mu) = \frac{\mu \delta(k-\tilde{k}_F)}{\vert{s v_F \hbar
\sqrt{\mu^2-(E_g/2)^2}}\vert}\\
\notag
\frac{\partial E_{\lambda}}{\partial k_{\alpha}} = 
\frac{s v_F^2 \hbar^2 k_{\alpha}}{\sqrt{\hbar^2 v^2_F k^2 +(E_g/2)^2}}\\
\notag
\frac{\partial^2 E_{\lambda}}{\partial k_{\alpha} \partial k_{\beta}} =
\frac{2 s v_F^2 \hbar^2 \left({E_g^2 \delta_{\alpha,\beta}+4 v_F^2 \hbar^2\left({k^2\delta_{\alpha,
\beta}-k_{\alpha} k_{\beta}}\right)}\right)}{\left({E_g^2+4 \hbar^2 v_F^2 k^2}\right)^{3/2}}, \
\end{gather}
with $s v_F \hbar \tilde{k}_F = \sqrt{\mu^2 -(E_g/2)^2}.$
Without loss of generality, we may assume $\mathbf{q} = (q_x,0)$ and $\mu>E_g/2>0$
so that $k_x = k \texttt{Cos}\phi$ and


\begin{equation}
\sum \limits_{\lambda} = \frac{4}{(2 \pi)^2} \int \limits_{0}^{\infty} k dk
\int \limits_{0}^{2 \pi} d \phi \ ,
\end{equation}
where the factor of four arises from the spin degeneracy. Making use of
 Eqs. \eqref{EQ:TRANSFORM} in Eq. \eqref{linearlong1}, a straightforward calculation
 shows that the  linear polarization function is given by

\begin{equation}
\Pi^{(1)}_{{\bf q},\omega}\approx\frac{q^2e^2\mu}{\pi\hbar^2\omega^2}
\left(1-\frac{E^2_g}{(2\mu)^2}\right)
\label{EQ:POLFORDER} \ .
\end{equation}
The polarization corresponding to SHG becomes 

\begin{equation}
\Pi^{(2)}_{2{\bf q},2\omega}=-\frac{3 e^3 q^4 v_F^2}{8 \pi \omega^4 \hbar^2}\left({1+3 (\frac{E_g}{2 
\mu})^2}\right)\left({1- (\frac{E_g}{2 \mu})^2}\right) \ .
\label{EQ:POLSORDER}
\end{equation}

\par
We now turn to calculating the observable  intensity of the SHG signal.
The part of the external filed running along the graphene sheet is characterized
by the potential

\begin{gather}
\Phi^{Ext}_{\mathbf{r},t} = \frac{\phi_0}{2}\texttt{Exp}\left({\mathbf{q r} - \omega t}\right) + c.c.\\
\notag
\Phi^{Ext}_{\mathbf{q},\omega} =  \frac{\phi_0}{2} \ .
\label{EQ:EXTERNAL1}
\end{gather}
Fourier transforming Poisson's equation for the induced field, we obtain

\begin{equation}
\Phi^{Ind}_{\mathbf{q},\omega}=\frac{2 \pi}{\epsilon_{\infty} q}\rho_{{\bf q},\omega}
=\frac{2 \pi}{\epsilon_\infty q}\Pi^{(1)}_{{\bf q},\omega}\Phi^{Tot}_{{\bf q},\omega} \ ,
\label{induced1}
\end{equation}
where $\epsilon_{\infty}$ is the dielectric constant of the substrate. On the other hand,

\begin{equation}
\Phi^{ind}_{{\bf q},\omega}=\Phi^{Tot}_{{\bf q},\omega}-\Phi^{Ext}_{{\bf q},\omega} \ .
\label{induced2}
\end{equation}
From the above two equations, we have

\begin{gather}
\notag
\Phi^{Tot}_{{\bf q},\omega}=\frac{\Phi^{Ext}_{{\bf q},\omega}}
{\epsilon({\bf q},\omega)}\\
\epsilon({\bf q},\omega)=1-\frac{2 \pi}{\epsilon_{\infty} q}\Pi^{(1)}_{{\bf q},\omega} \ .
\label{EQ:SPECTRALLIN}
\end{gather}

\par

The plasmon resonances are given by the solutions of $\epsilon({\bf q},\omega)=0$.
By  using the Drude formula for the dielectric function, i.e.,

\begin{equation}
\epsilon({\bf q},\omega)=1-\frac{\omega^2_p}{\omega^2} \ ,
\end{equation}
we obtain the plasmon dispersion relation

\begin{equation}
\omega^2_p=\omega_0^2 \left(1-\frac{E^2_g}{(2\mu)^2}\right) \ ,
\end{equation}
where $\omega_0^2 = 2 e^2 \mu q/ \epsilon_{\infty} \hbar^2$. We may also
 introduce the dimensionless quantity $(\omega_0/\mu)^2 = (2.5/\epsilon_{\infty}) (q/k_F)$.
This agrees with our previous calculations\cite{roslyak}. When second-order corrections
are included in the solution of Poisson's equation, our calculation shows that
\begin{eqnarray}
\Phi^{Ind}_{{\bf r},t}&=&\frac{2\pi}{q}
\big[\Pi^{(1)}_{{\bf q},\omega}\Phi^{Tot}_{{\bf q},\omega}
e^{i({\bf q\cdot r}-\omega t)}\nonumber\\
&+&\frac{1}{2}\Pi^{(1)}_{2{\bf q},2\omega}\Phi^{Tot}_{2{\bf q},2\omega}
e^{2i({\bf q\cdot r}-\omega t)}\nonumber\\
&+&\frac{1}{2}\Pi^{(2)}_{2{\bf q},2\omega}\Phi^{Tot}_{{\bf q},\omega}
\Phi^{Tot}_{{\bf q},\omega}+\;\;\text{c.c.}\big]\nonumber\\
&=&\Phi^{ind}_{{\bf q},\omega}e^{i({\bf q\cdot r}-\omega t)}
+\Phi^{ind}_{2{\bf q},2\omega}e^{2i({\bf q\cdot r}-\omega t)}
+\;\;\text{c.c.} \ .
\end{eqnarray}
\par
Taking  into account the fact that we have

\begin{eqnarray*}
\Phi^{Tot}_{{\bf q},\omega}&=&\Phi^{Ext}_{{\bf q},\omega}
+\Phi^{ind}_{{\bf q},\omega}\\
\Phi^{Tot}_{2{\bf q},2\omega}&=&\Phi^{Ext}_{2{\bf q},2\omega}
+\Phi^{ind}_{2{\bf q},2\omega} \ ,
\end{eqnarray*}
we arrive at  two regimes which are

\begin{enumerate}

\item Narrow-band perturbation satisfying   $\Phi^{Ext}_{2{\bf q},2\omega}
\approx 0$  yields

\begin{eqnarray}
\Phi^{Tot}_{2{\bf q},2\omega}&=&\Phi^{Ind}_{2{\bf q},2\omega}
=\frac{\pi}{q}\frac{\Pi^{(2)}_{2{\bf q},2\omega}}
{\epsilon(2{\bf q},2\omega)}\Phi^{Tot}_{{\bf q},\omega}\Phi^{Tot}_{{\bf q},\omega}
\nonumber\\
&=&\frac{\pi}{q}\frac{\Pi^{(2)}_{2{\bf q},2\omega}}
{\epsilon(2{\bf q},2\omega)\epsilon({\bf q},\omega)}
\Phi^{Ext}_{{\bf q},\omega}\Phi^{Ext}_{{\bf q},\omega} \ .
\label{plasmon}
\end{eqnarray}

\item In the broad-band limit, $\Phi^{Ext}_{2{\bf q},2\omega}\approx
\Phi^{Ext}_{{\bf q},\omega} $, yielding

\begin{eqnarray}
\Phi^{Tot}_{2{\bf q},2\omega}&=&
\frac{1}{\epsilon(2{\bf q},2\omega)}\Phi^{Ext}_{{\bf q},\omega}\nonumber\\
&+&\frac{\pi}{q}\frac{\Pi^{(2)}_{2{\bf q},2\omega}}{\epsilon(2{\bf q},2\omega)\epsilon^2({\bf q},\omega)}
\Phi^{Ext}_{{\bf q},\omega}\Phi^{Ext}_{{\bf q},\omega} \nonumber\\
&\approx&
\frac{1}{\epsilon(2{\bf q},2\omega)}\Phi^{Ext}_{{\bf q},\omega}.
\label{EQ:BROADBAND}
\end{eqnarray}

\end{enumerate}

\par
Owing to the linear dependence on $\Phi^{Ext}_{{\bf q},\omega}$, the broad-band
signal is usually dominated by linear absorption. Consequently,  we concentrate
our attention  on the first case.  The poles of Eqs. \eqref{plasmon} and
\eqref{EQ:BROADBAND} correspond to  the new plasmon modes. The double resonance
condition $\omega^2(2q)=2\omega^2(q)$  never occurs in the long wavelength
regime, which means that  we have two separate plasmon branches at $\omega=\omega_p$ and 
$\omega=\omega_p/\sqrt{2}$.

\par
The total intensity of the  measured and external fields is given by

\begin{gather}
I^{Tot}_{2{\bf q},2\omega}=\frac{c}{8\pi}\nabla\Phi^{Tot}_{2{\bf q},2\omega}
\cdot\nabla\Phi^{Tot}_{2{\bf q},2\omega}\\
\notag
I^{ext}_{{\bf q},\omega}=\frac{c}{8\pi}\nabla\Phi^{Ext}_{{\bf q},\omega}
\cdot\nabla\Phi^{Ext}_{{\bf q},\omega}
\label{EQ:INTENSITY}
\end{gather}
From this equation as well as  Eqs. \eqref{induced2} and \eqref{EQ:POLSORDER},
 we finally obtain the normalized SHG  intensity given by
\begin{widetext}
\begin{eqnarray}
\frac{I^{Tot}_{2{\bf q},2\omega}}{\left({I^{Ext}_{{\bf q},\omega}}\right)^2}&=&
\frac{9 \pi e^6 q^4 v^4_{F} \left({\gamma^2 + \omega^2}\right)^2
\left({\gamma^2 +4 \omega^2}\right)^2 \left({1-(E_g/2 \mu)^2}\right)^2  
\left({1+3(E_g/2 \mu)^2}\right)^2}
{2 c \omega^2 \hbar^4 \left({\gamma^2 \omega^2 + \left({\omega^2 -\omega_p^2}\right)^2}\right)^2 
\left({\gamma^2 \omega^2 + \left({2 \omega^2 -\omega_p^2}\right)^2}\right)}  \ .
\label{secondharmoic}
\end{eqnarray}
\end{widetext}
This  expression is the main result of our paper and will be discussed in the following
 section.
 
\begin{figure}[!h]
\centering
\includegraphics[width=0.48\textwidth]{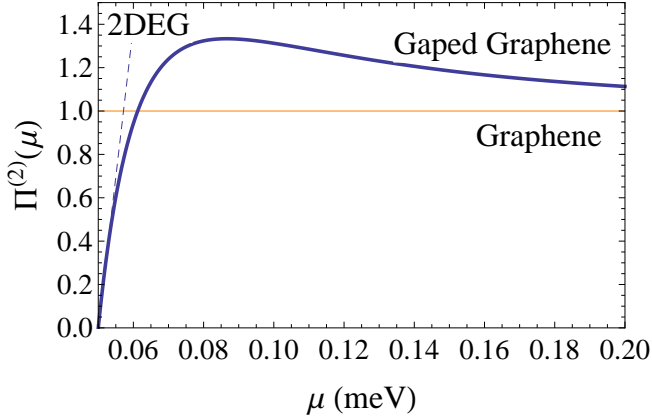}
\caption{(Color online) The 2DEG polarization (in units of $\Pi^{(2)}(E_g=0)$ as a function of electro-chemical potential for chosen $E_g=11.6 \; meV$}
\label{FIG:5}
\end{figure}
\begin{figure}[!h]
\centering
\includegraphics[width=0.48\textwidth]{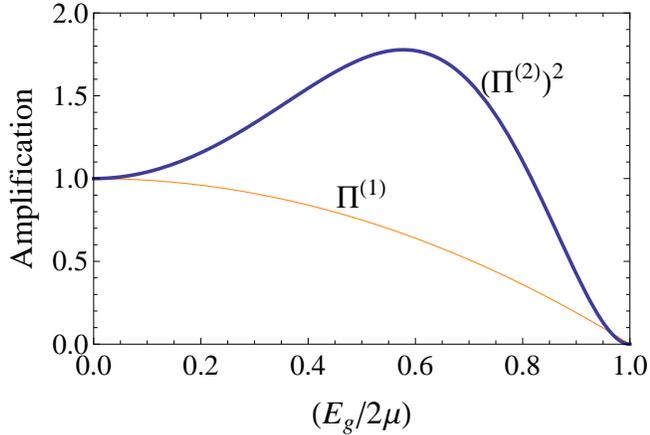}
\caption{(Color online) Plots of the polarization functions in  \eqref{EQ:POLFORDER} 
and \eqref{EQ:POLSORDER} with $E_g/2\mu$ in units of their value for $E_g=0$}
\end{figure}
\begin{figure}[h]
\centering
\includegraphics[width=0.8\textwidth]{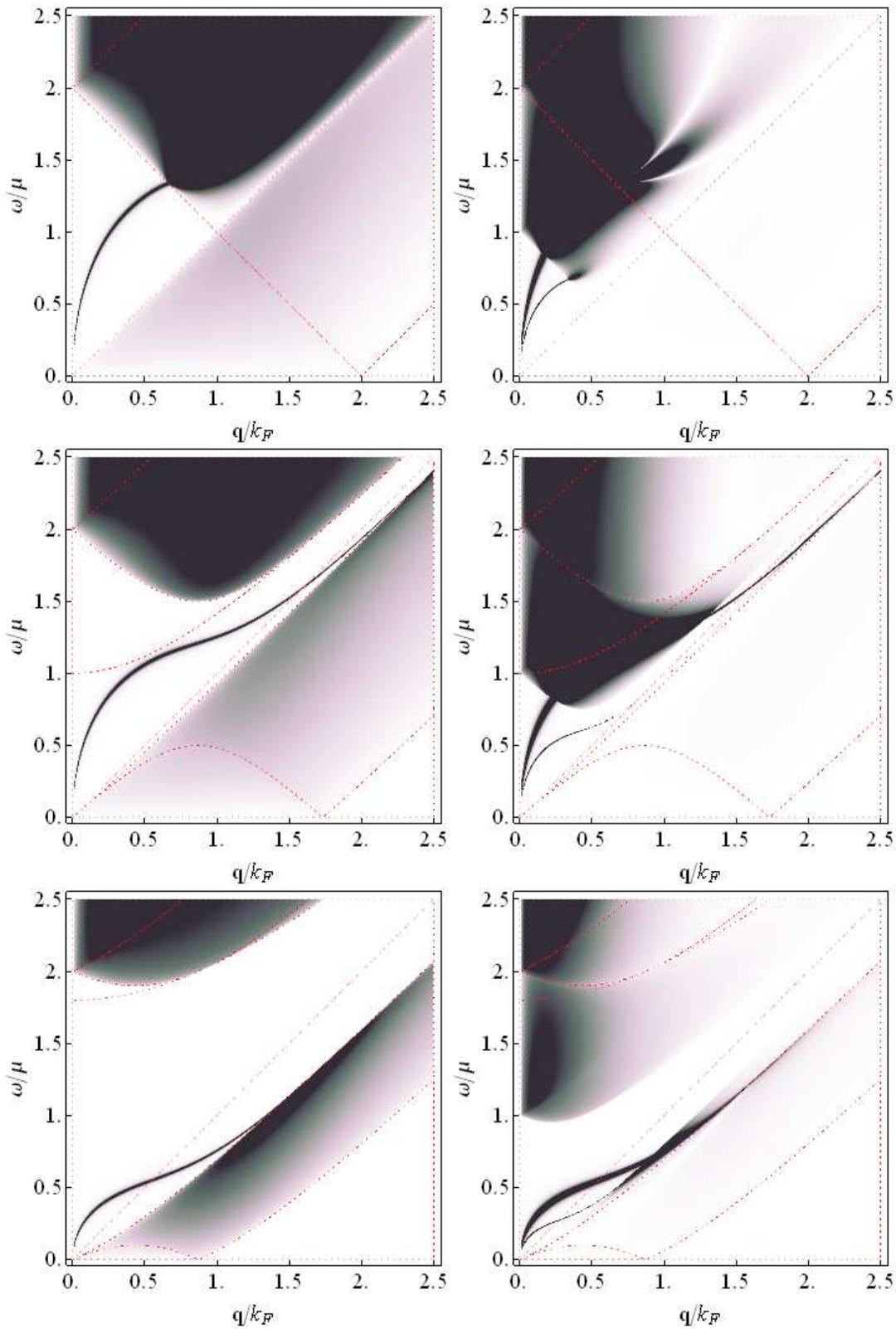}
\caption{(Color online) Poles of the imaginary part of the spectral function. 
The left/right  panels are density plots for linear absorption/SHG, respectively.
For concreteness, the plasmon dephasing is chosen as    $\gamma/2\mu = 0.01.$}
\label{FIG:4}
\end{figure}
\begin{figure}[!h]
\centering
\includegraphics[width=1.0\textwidth]{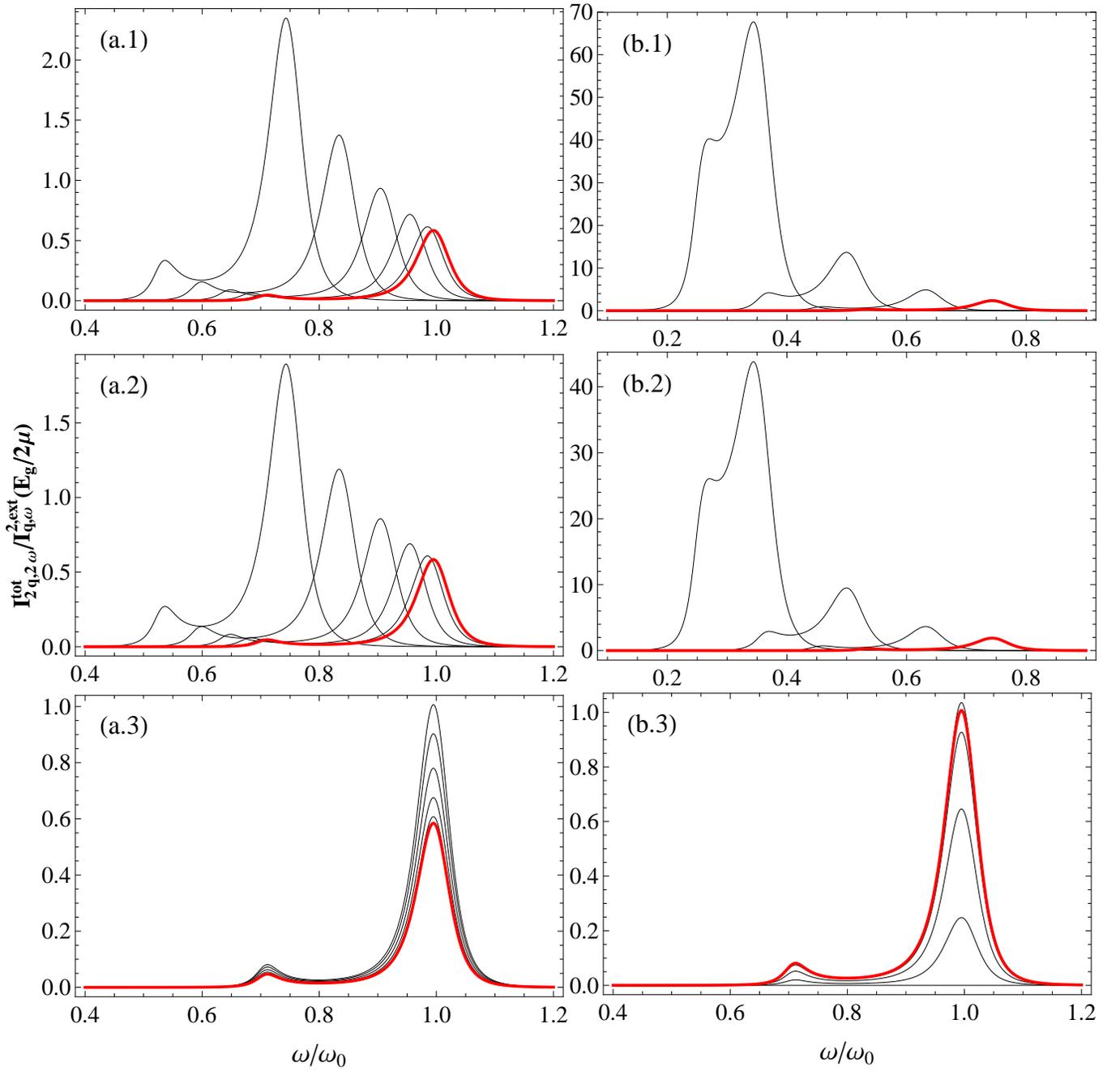}
\caption{(Color online) Left panels show an increase starting with  $E_g/2\mu=0.0$  (red curve),
then  $E_g/2\mu=0.1,\cdots.,0.5$. Right panels show growth for  $E_g/2\mu=0.5$ (red curve), and 
$E_g/2\mu=0.6,\cdots,0.9$. Panels 2 are the effect due to change in $\omega_p$ only, 
The plasmon dephasing is set at $\gamma/2\mu = 0.1$.}
\label{FIG:6}
\end{figure}
\medskip
\par
\begin{figure}[!h]
\centering
\includegraphics[width=0.48\textwidth]{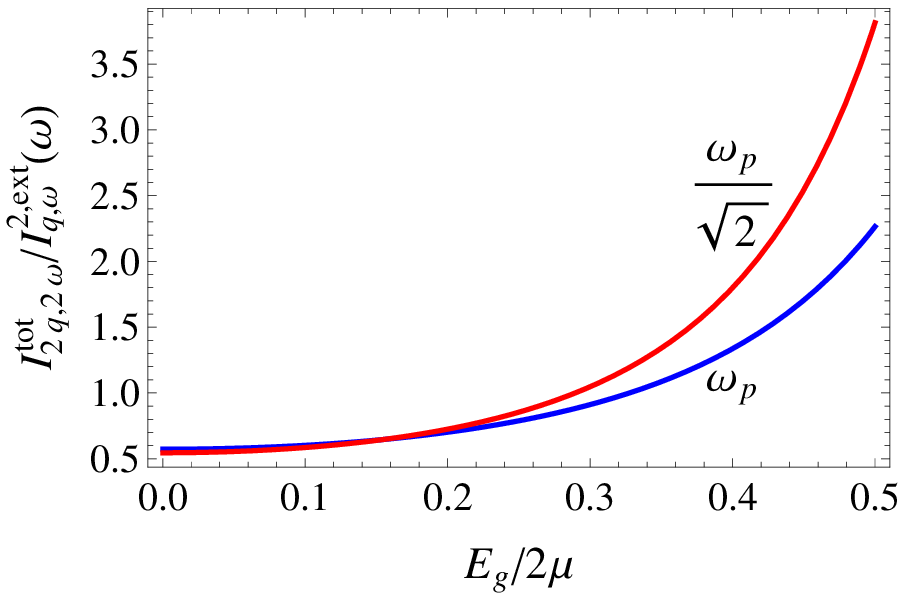}
\caption{(Color online) Variation of the SHG signal along $\omega_p$ and $\omega_p/\sqrt{2}$. 
The latter was scaled up by a factor of twelve, for convenience.}
\label{FIG:7}
\end{figure}


\section{Numerical Results and Discussion}

We now complement our formalism by numerical simulations.
For concreteness, we  assume that the  wavelength of incoming light to be
$\lambda = 800 \; nm$  and the  angle of incidence measured from  the normal to the surface
is $45^\circ$,  as shown schematically in Fig. \ref{FIG:2}. The chemical potential is
fixed by choosing  $q/k_F =0.1$, thereby making $k_F= 8.83 \times 10^6\; m^{-1}$,
i.e.,  $\mu = 5.8 \; meV$. This value of the chemical potential is well within the
Dirac cone approximation. Consequently, we have $\omega_p (E_g=0)=\omega_0 =1.54 \; meV$,
and $\omega_0/\sqrt{2}=1.09 \; meV$. Those values are much smaller than the incoming
light energy of $1.55 \; eV$. This indicates that graphene is mostly transparent to
light and only a small portion of it is specularly reflected even in the linear regime.
The gap-inducing substrate is taken to be boron-nitride ($\texttt{BN}$) with
background dielectric constant $\epsilon_{\infty} = 7.1$. The nitrogen atoms of the
substrate are in the center of the carbon-formed hexagons. In Ref. \cite{giovannetti}, it
was shown that the induced gap   depends inversely on the distance $d$ from the graphene
layer so  that $d=3.65 \; \AA$ corresponds to $E_g/2 \mu = 1$.

\par
Some useful information regarding the SHG signal may be directly extracted  from the 
poles of the spectral function in Eqs. \eqref{EQ:SPECTRALLIN} and \eqref{plasmon} without 
employing  the long wavelength approximation. In the left-hand  column of Fig. \ref{FIG:4}, 
the linear response of  gapped graphene is presented. Clearly, there is  cross-over from 
Dirac  ($E_g/2 \mu=0$) to 2DEG-like ($E_g/2\mu=0.9$) plasmon behavior. The SHG possess   
poles as shown in the right-hand panel of Fig. \ref{FIG:4}. Of the two plasmon branches, the 
one at $\omega_p(q)$ is suppressed by  linear response, whereas the one at $\omega_p(q)/\sqrt{2}$ 
may be spectrally  resolved. However, when $E_g/2 \mu = 0$, both branches   are Landau 
damped when $\omega/\mu>1-q/k_F$.  Once the gap is increased to $E_g/2 \mu =0.5$, the 
lower branch may appear in a region which opens up within the electron-hole continuum 
 and is  undamped  beyond the long wavelength limit.  For larger values of the gap, both 
 branches merge  with the electron-hole continuum at the same value of $q/k_F$. As the 
 gap is further increased,  both plasmon  frequencies are reduced in accordance with the 
 reduction in the linear  response polarization function, as indicated in the 
 right panel of Fig.\ref{FIG:5}.
Consequently, the spectral separation between them gets reduced, thereby making it 
 more difficult to detect the lower SHG branch.

\par
In order to study relative intensities of these plasmon branches, we must resort to the 
full version of the intensity ratio given in Eq.  \eqref{secondharmoic}, thereby 
limiting ourselves to the long wavelength regime. One of the main factors determining 
that ratio is the square of the second-order polarization shown in Fig. \ref{FIG:5}.
When $E_g/2\mu=0.6$, the second -order polarization reaches its maximum value which is 
seventy times larger than that of gapless graphene. To explain the  maximum, it is convenient 
to fix the value of the gap at $E_g = 11.6 \; meV$ and then vary the chemical potential as 
shown in the left panel of Fig. \ref{FIG:5}. For small values of the chemical potential we have 2DEG-like 
behavior with the second-order polarization $\sim \mu$. For its large values, we have Dirac-like
 behavior with the second order polarization being independent of the chemical potential.
Therefore, the maximum is the cross-over point between those two regimes.

\par
The experimentally measurable Eq.\eqref{secondharmoic} before and after cross-over is
shown in panels (a) and (b) of Fig. \ref{FIG:6}. As we mentioned above, there are two 
factors affecting SHG intensity: the second-order polarization, given by the numerator, and the change 
in the plasmon frequency, given by the denominator, in Eq.\eqref{secondharmoic}).
Their separate effects are shown in panels (2) and (3) of Fig. \ref{FIG:6}.
Those two effects work in favor of each other  before the cross-over and against each other after 
that. Nevertheless, we observe steady growth of SHG intensity with $E_g$, making it 
an order of magnitude larger than that of conventional graphene. Fig.\ref{FIG:7} 
demonstrates that the lower $\omega_p/\sqrt{2}$ plasmon branch continues to grow
 with increased $E_g/2 \mu$. This opens up an experimental avenue to identify those 
 branches without relying on their spectral separation. This is similar to the effect 
 of DC current on SHG but without underlying anisotropy induced by the current\cite{bykov}.


\section{Concluding Remarks}

We have investigated the influence of substrate-induced gap in graphene on SHG signal.
The maximum of the signal was attributed to an additional plasmon branch at $\omega_p/\sqrt{2}$.
A red shift and an order of magnitude enhancement of that resonance with 
 increased gap or reduced electro-chemical potential was demonstrated. The intensity
  of that branch increases more rapidly than the conventional $\omega_p$ branch which 
  compensates for their reduced spectral separation. Our formalism is an alternative 
  to DC induced enhancement in SHG but without accompanying the latter anisotropy in SHG signal.

\begin{acknowledgments}
This research was supported by contract \# FA 9453-07-C-0207 of AFRL. 
\end{acknowledgments}


\begin{thebibliography}{9}
\bibitem{offset} For In$_x$Ga$_{1-x}$As, the energy gap $E_g$ (in eV) is
$1.422-1.53x+0.45x^2$; for Al$_x$Ga$_{1-x}$As, the energy gap (in eV) is
$1.422(1-x)+2.677x$. The band offset for the conduction band is taken to
be $\Delta E_c=0.70\ \Delta E_g$ and $\Delta E_v=0.30\ \Delta E_g$
for the valence band, where $ \Delta E_g$ is the discontinuity
of the
energy gap for the two bulk materials forming the adjacent layers.
\end{thebibliography}
\end{document}